\documentclass[aps,prb,twocolumn,groupedaddress,showpacs]{revtex4}

\usepackage{longtable}
\usepackage{graphicx}
\usepackage{dcolumn}
\usepackage{bm}
\usepackage{amsmath}
\usepackage[psamsfonts]{amssymb}

\newcommand{\MR}{magnetoresistance}

\newcommand{\etal}{\emph{et al.}}

\begin{document}

\title{Phase coherent transport in SrTiO$_3$$/$LaAlO$_3$ interfaces}


\author{D. Rakhmilevitch}
\affiliation{Raymond and Beverly Sackler School of Physics and
Astronomy, Tel-Aviv University, Tel Aviv, 69978, Israel}
\author{M. Ben Shalom}
\affiliation{Raymond and Beverly Sackler School of Physics and
Astronomy, Tel-Aviv University, Tel Aviv, 69978, Israel}
\author{M. Eshkol}
\affiliation{Raymond and Beverly Sackler School of Physics
and Astronomy, Tel-Aviv University, Tel Aviv, 69978,
Israel}
\author{A. Tsukernik}
\affiliation{Raymond and Beverly Sackler School of Physics
and Astronomy, Tel-Aviv University, Tel Aviv, 69978,
Israel}
\author{A. Palevski}
\affiliation{Raymond and Beverly Sackler School of Physics and
Astronomy, Tel-Aviv University, Tel Aviv, 69978, Israel}
\author{Y. Dagan}
\email[]{yodagan@post.tau.ac.il} \affiliation{Raymond and Beverly
Sackler School of Physics and Astronomy, Tel-Aviv University, Tel
Aviv, 69978, Israel}

\date{\today}
\begin{abstract}
The two dimensional electron gas formed between the two band
insulators SrTiO$_3$ and LaAlO$_3$ exhibits a variety
of interesting physical properties which make it an appealing
material for use in future spintronics and/or quantum computing
devices. For this kind of applications electrons have to retain
their phase memory for sufficiently long times or length. Using a
mesoscopic size device we were able to extract the phase coherence
length, $L_\phi$ and its temperature variation. We find the dephasing
rate to have a power law dependence on temperature. The power depends on the temperature range studied and sheet resistance as expected from dephasing due to strong
electron-electron interactions.
\end{abstract}


\maketitle
Oxides interfaces have recently attracted considerable scientific
attention due to their potential for implementation in oxide based
electronics and for basic science. One of the most studied
examples is the interface between LaAlO$_3$ and SrTiO$_3$
(LAO/STO).\cite{OhtomoHwang} In this interface a highly conducting two
dimensional electron gas (2DEG) appears above a critical thickness
of 4 unit cells.\cite{Thiel}
\par
Magnetotransport in the 2DEG formed at the interface between STO and LAO
exhibits quite rich behavior in low and high magnetic fields in
both perpendicular and parallel orientations.\cite{benshalom} The magnetoresistance at low magnetic fields
was described in terms of suppression of weak localization.\cite{CavigliaSpinorbit}
However, it is difficult to sort out the
contribution of quantum corrections to the \MR~ from other possible mechanisms such as two types of charge carriers and inhomogeneities.There is therefore a need for
a different type of measurement sensitive solely to the mesoscopic phase
coherence, such as universal conductance fluctuations (UCF).
\cite{Lee_and_Stone,UCF_in_metals,UCF_in_semiconductors}
It has been shown that the conductivity of sample with a given
impurity concentration will fluctuate around it's average value
upon changing the impurity configuration. For two dimensions these
fluctuations should have a universal value of the order of $\frac{e^2}{h}$.
\cite{Lee_and_Stone} An experimental realization simulating such a
configuration modification is achieved by an application of
external magnetic field.\cite{Ergodic_PhysRevLett.55.1622}
In this case the magneto-conductance will produce a reproducible
fluctuating curve. In order to observe this effect the size of the
sample should be comparable with the low temperature length
scales: $L_\phi$, the phase breaking length and/or L$_T=\sqrt{\frac{hD}{k_BT}}$ the
so-called thermal length. The root mean square (RMS) value of the UCF and the width
of the typical oscillation provide information on L$_T$ and
L$_\phi$ when the length and the width of the sample exceed the
above microscopic lengths\cite{UCF_finite_temperatures, Thornton}.
\par
For a macroscopic sample the UCF average out. However, coherent
back scattering also known as weak localization (WL)\cite{AbrahamsScalingAALR, AltshulerMR, BergmanWLSO} still affects the
conductance. The latter quantum corrections as well as the UCF
depend on the same microscopic length-scale, $L_\phi$. The
observation of both the WL and UCF with a similar $L_\phi$ gives a
strong evidence that the low field \MR~ in macroscopic samples is
governed by quantum interference effects.
\par
In this paper, we report studies of magnetoconductance
fluctuations in mesoscopic samples of the 2DEG at the STO/LAO
interface as well as WL in a macroscopic sample. The analysis of the UCF suggests that the low field
\MR~ for macroscopic samples at the low carrier densities is
governed by quantum corrections, namely weak localization. In
addition, we show that for these densities the dephasing mechanism
is mainly due to electron-electron interactions.
\par
\begin{figure}
\includegraphics[width=1.1\hsize]{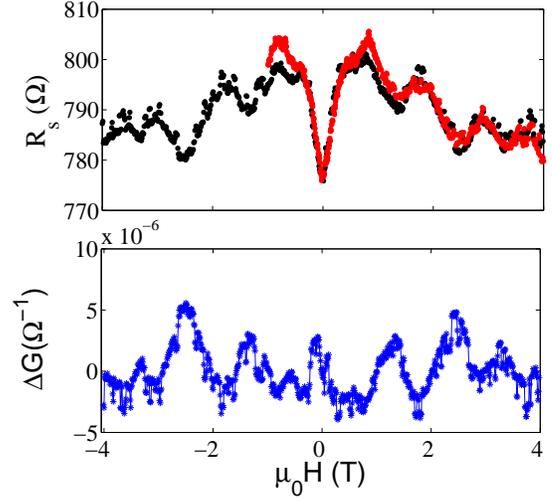}
\caption {(color on-line) a. Sheet resistance versus magnetic field of the 8uc bridge at 1.3 K. Two traces of the magnetic field sweep are shown (red circles and black squares) note the reproducible fluctuations pattern .b. The deviation of the magnetoconductance from its average value (see text for details).\label{fluctuations}}
\end{figure}
8 and 20 unit cell (uc) samples of LAO were epitaxially grown on
atomically flat TiO$_2$ terminated STO substrate using pulsed
laser deposition. The growth parameters are described
elsewhere.\cite{benshalom} The 8uc sample was patterned into
a Hall-bar geometry using a combination of photo- and
electron-beam lithography. The LAO was dry-etched using Ar ions.
The etching was followed by a short oxygen treatment at 200$^\circ
C$ in order to anneal out oxygen vacancies created in the substrate during the
etching. The dimensions of the Hall bar are $4 \mu$m$\times 1.5
\mu$m. The 20uc sample was measured
in a Van der Pauw configuration. The samples were measured in
$^4$He cryostat (8uc) and $^3$He refrigerator (20uc)
\par

\begin{figure}
\includegraphics[width=1.1\hsize]{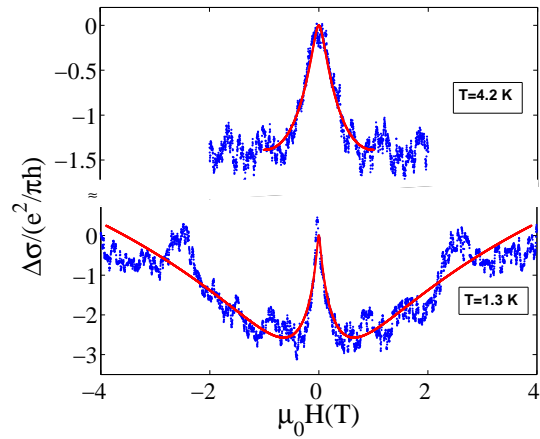}
\caption {(color on-line) Magnetoconductance at 1.3K and at 4.2K for the 8uc sample. The solid lines are fits to equation(5).\label{localization}}
\end{figure}

The typical magnetoresistance curve of the bridge is shown in Fig.\ref{fluctuations}a. The field
was cycled from -4 to 4 T for several times at 1.3 K and at 4.2 K.
Two of these traces at 1.3 K are shown. A reproducible fluctuating
signal is clearly seen. These are the universal quantum
fluctuations. Fig.\ref{fluctuations}.b shows conductance deviations from its average
as a function of field. The subtracted average background
magneto-conductance is obtained by removal of the WL fit plotted in Fig.\ref{localization} .
\par
The conductance-correlation function:
\begin{equation}
	F(\Delta B)=<\delta g(B)\delta g(B+\Delta(B)>
\end{equation}

was calculated from the data in Fig.\ref{fluctuations}b. $F(\Delta B=0)$ gives the RMS
value of the fluctuations. The magnetic correlation field $B_c$ is
found from
\begin{equation}
	F(\Delta B=B_c)=\frac{1}{2}F(\Delta B=0)
\end{equation}

For sample dimensions L with $L > L_\phi \approx L_T$ the RMS value is given by\cite{UCF_SO, UCF_valley, Spivak}
\begin{equation}
\delta G=\frac{N_v\alpha}{\beta}\cdot\frac{e^2}{h}\cdot\sqrt{\frac{4W}{\pi L^3}}\cdot L_{\phi}
\end{equation}

where $\beta$ represents the effect of SO coupling on the size of the fluctuations. $\beta$ goes to 2 at the strong coupling limit.\cite{UCF_SO} The $N_v\alpha$ factor stands for the effect of valley degeneracy and is the same as the factor that appears in weak localization theory.\cite{UCF_valley} $\alpha$ has been theoretically predicted to be of order 1. From this relation using our measured RMS value and the known sample dimensions we find $L_\phi=1570\pm300$ \AA \ \ at 1.3 K. Here we assume that spin-orbit scattering length, L$_{so}$ is much shorter than the dephasing
length L$_\phi$ and therefore $\beta\approx2$. This is reasonable in view of the strong spin-orbit coupling
observed in this material.\cite{BenSHalomPRL} The valley degeneracy $N_v$ is assumed to be 3 as in bulk STO.\cite{STO_SdH} The consistency of this assumption as well as the assumption $L_\phi \approx L_{T}$ will be addressed further below.
\par
The dependence of $B_c$ on $L_{\phi}$ has been calculated,\cite{Spivak}

\begin{equation}
	 L_{\phi}=\sqrt{\frac{h/e}{B_c}}
\end{equation}

For $L_\phi \approx L_{T}$. Substituting the obtained value for $B_c$ we find
L$_\phi=1650\pm320$ \AA, consistent with the value estimated from the RMS value of the fluctuation amplitude.
\par

The magnetoconductance dip at zero magnetic field was analyzed in the
framework of WL theory in the presence of spin-orbit interactions.\cite{HikamiSOMR, WL_MF_fit} The data and the theoretical curves using equation (5) are presented in fig.\ref{localization}
 \[
\Delta\sigma(H,T)_{WL}=\frac{e^2}{\pi h}N_v\alpha[\Psi(x_1)+ \frac{1}{2\sqrt{1-\gamma^2}}\Psi(x_2)
\]
\begin{equation}
-\frac{1}{2\sqrt{1-\gamma^2}}\Psi(x_3)]
\end{equation}
where
\[
\Psi(x)=ln(x)+\psi(\frac{1}{2}+\frac{1}{x}), x_1=\frac{H}{H_i+H_{so}},
\]
\[
 x_2=\frac{H}{H_i+H_{so}(1+\sqrt{1-\gamma^2})}, x_3=\frac{H}{H_i+H_{so}(1-\sqrt{1-\gamma^2})},
\]
$\gamma=g\mu H/(4eDH_{so})$ and $N_v=3$.\cite{STO_SdH} We used $\alpha\approx 1$ as theoretically predicted for inter-valley scattering rate smaller than the dephasing one\cite{WL_valley_theory} and $g=2.25$ consistent with Ref.\cite{CavigliaSpinorbit} for sample with a similar sheet resistance.
L$_\phi(T)$ and L$_{so}$ are used as the fitting parameters.\cite{remark2} From the fit we find
L$_\phi=1300\pm260$ \AA~ consistent with the values found from UCF and L$_{so}\approx200$ \AA~. It is indeed significantly shorter than $L_\phi$ at 1.3 K as we assumed earlier in the analysis of the UCF data.
It is important to stress that the value of L$_\phi$ deduced from the fit
to the WL calculations is similar to that extracted from the UCF. This provides an important evidence that weak localization corrections play an important role in the low field magneto-transport.
\par
At 4.2K $L_\phi=640\pm100$ \AA. It is much shorter than at 1.3 K. We note that the
ratio of the diffusion coefficient for these temperatures is
$\frac{D(4.2 K)}{D(1.3 K)}=\frac{R(1.3 K)}{R(4.2 K)}=1.5$. Since
L$_\phi=\sqrt{D \tau_\phi}$ we obtain $\frac{\tau_\phi(4.2
K)}{\tau_\phi(1.3 K)}=6$, much higher than the ratio between the
phase coherence lengths. In order to elucidate the mechanism responsible for dephasing we
carefully studied the temperature dependence of the \MR~. We used
the macroscopic 20uc sample since in such a sample the analysis of
the central weak localization feature is less affected by the UCF themselves, which can alter the conductance at low fields. In principle, one could study a mesoscopic sample but for our case the number of
fluctuations is too small to provide an accurate measurement for the temperature dependence of L$_\phi$.
\par
\begin{figure}
\includegraphics[width=1\hsize]{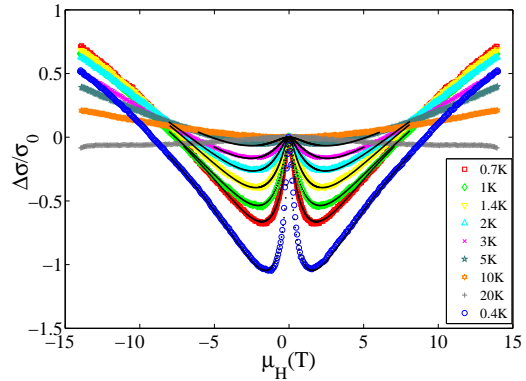}
\caption {(color on-line) The normalized magneto-conductance for the 20uc sample at various temperatures. The solid lines are fits to equation (5).\label{allmr}}
\end{figure}

In fig.\ref{allmr} we present the magnetoconductance curves for the 20uc sample for temperatures ranging from
0.35 K to 20 K. The solid lines are the theoretical fits using equation (5)
Here we also used L$_\phi(T)$ and L$_{so}$ as the fitting parameters. The data was fit for a field range between -5 T to  5 T (or more). The higher field region was not included in the fit to avoid high
field contributions due to electron-electron interactions and
other possible corrections.\cite{RevModPhys} The fit is excellent for T$<5$K where quantum corrections dominate the low field behavior. For this sample we used $g=0.6\pm0.1$ (consistent with Ref.\cite{CavigliaSpinorbit} for similar sheet resistance). We allowed g to vary within the error bar for the various temperatures to obtain a better fit for the higher field regime. This small variation does not effect the value of L$_\phi$.
The data could not be fit with $N_v\alpha=3$ but with $N_v\alpha=1$. The diminished role of the valley degeneracy for this sample can be explained by a stronger inter-valley scattering.\cite{WL_valley_exp, WL_valley_theory} We should keep in mind that this sample has mobility $\approx$6 times lower than the 8uc sample. In addition, $\alpha N_v=1$ was also used in Ref.\cite{CavigliaSpinorbit} for samples with similar sheet resistance.
We note that while the magnetoresistance can be well explained by equation (5) the temperature dependence of the resistivity is more complicated. Other effects not included in our simple analysis may be involved. However, since the localization peak is seen together with the UCF we can be sure that quantum corrections are dominate the magneto-conductance at low temperatures.

\par
The quasi-elastic electron-electron scattering time is known to be dependent on disorder. At low temperatures, in most cases, this is the major mechanism for the dephasing. This can explain the difference in dephasing between the two samples. To verify this assumption we plot the dephasing rate as function of temperature for both samples.

\begin{figure}
\includegraphics[width=1\hsize]{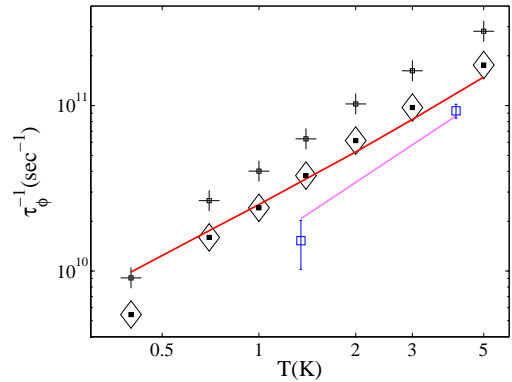}
\caption {(color on-line) The dephasing rate extracted from the fits in Figs.\ref{localization} and \ref{allmr} for and 20uc samples using $m^\ast=3m_e$ (crosses) $m^\ast=5m_e$ (diamonds) and for 8uc with $m^\ast=3m_e$ (squares). The lines are the theoretical curve using equation (6).\label{times}}
\end{figure}

In fig.\ref{times} the dephasing rate calculated from L$_\phi=\sqrt{D\tau_\phi}$ is plotted as a function of temperature on a logarithmic scale for both samples. The diffusion coefficient was calculated from Einstein relation $D=\frac{\pi\hbar^2}{N_vm^\ast e^2 R_\square}$. We used an effective mass of $m^\ast=3m_e$ (m$_e$ the bare electron mass) and $N_v=3$ as found for bulk SrTiO$_3$. A liner temperature dependence is observed for the 20uc sample. The dephasing rate extrapolates to zero at zero temperature. This temperature dependence is expected from a Nyquist noise resulting from electron-electron interactions. For the 8uc sample a higher exponent is observed. Narozhny \etal.\cite{Alainer} calculated the dephasing rate for this case and found it to depend on the sheet resistance. This theory well describes the dephasing rate in semiconductors.\cite{Eshkolpaper}

\[
\frac{1}{\tau_{\phi}}=\left\{1+\frac{3(F_0^{\sigma})^2}{(1+F_0^{\sigma})(2+F_0^{\sigma})}\right\}\frac{k_BT}{\eta\hbar}ln[\eta(1+F_0^{\sigma})]
\]
\begin{equation}
+\frac{\pi}{4}\left\{1+\frac{3(F_0^{\sigma})^2}{(1+F_0^{\sigma})^2}\right\}\frac{(k_BT)^2}{\hbar E_F}ln(E_F\tau/\hbar)
\end{equation}

where $F_0^{\sigma}$ is the interaction constant in the triplet channel, which depends on interaction strength,\cite{ZalaFsigmaTripletchannel}, $\eta = \frac{2 \pi \hbar}{e^2 R_\square}$, $k_B$ is the Boltzmann constant and $E_F$ is the Fermi energy.
\par
The solid lines in Fig.\ref{times} are theoretical curves for the dephasing rate using the measured parameters for our samples. We used $F_0^{\sigma}=0$ for both samples since the interaction parameter r$_s$ is expected to be small. The dephasing rate calculated from the 20 uc data fits the theoretical curve using m$^\ast= 5 m_e$\cite{remark} we also show $\frac{1}{\tau_\phi}$ for m$^\ast =3 m_e$. Since the theoretical calculation is accurate within a factor of 2 one can safely state that both sets of $\tau_\phi ^{-1}$ agree with the theoretical curve. The dephasing rate for the 8 uc sample at 1.3 K $\tau_phi ^{-1}\approx 2.1\times10^{10}$ sec$^-1$ which is indeed very close to $\frac{k_B T}{h} \approx 2.7 \times10^{10}$ sec$^-1$ thus justifying the assumption of $L_\phi \approx L_T$ use in the UCF analysis.
\par
For the 20uc sample the sheet resistance is high, therefore the first term in equation(6) dominates. This term gives the observed linear temperature dependence. For the 8uc sample the sheet resistance is 3 times lower. Consequently the quadratic term has a significant contribution for the temperature range studied.
The fact that data fit the theoretical curves put strong limitations on the product N$_v m^\ast$, which appears in the diffusion coefficient. Moreover, it can be seen in equation(6) that for low resistance samples the temperature dependence of the dephasing rate depends on the Fermi energy and hence on the the product N$_v m^{\ast}$ ($\alpha$ does not enter into the calculation of $\tau_\phi$). The importance of the quadratic term in our data (8uc) can therefore provide additional information on the effective mass. Hence we can conclude that N$_vm^\ast$ should be very close to 9m$_e$. If we assume that $m^\ast$ is very large (say $m^\ast=9$ and N$_v$=1) it will be difficult to explain the absence of a quadratic term in the dephasing rate for the 20uc sample. If $m^\ast$ is small the measured dephasing rate will go below the theoretical limit set by the Nyquist noise. We therefore conclude that $m^\ast$ is of the order of a few $m_e$. Similar results were obtained from infrared ellipsometry.\cite{Bernhard_elipsometry}
\par

In summary, we measured universal conductance fluctuations in a mesoscopic sample of LaAlO$_3/$SrTiO$_3$ interface. The phase coherence length obtained from these fluctuations equals (within the error) to that found from fitting the low field magnetoresistance curves. This suggests that the \MR~ at low fields is dominated by quantum corrections. We analyze the temperature dependence of the \MR~ in terms of the week localization theory. From the temperature dependence of the dephasing rate we conclude that it is dominated by electron-electron interactions. The dephasing time fits very well to the theoretical calculations of Narozhny \etal.\cite{Alainer} assuming an effective mass of 3$m_e$. Since dephasing time is only limited by electron-electron interactions it seems that increasing the carrier concentration by gate voltage application will enhance the phase coherent length over the micrometer scale, meeting the requirements of quantum coherent electronic devices.

\begin{acknowledgments}
This research was supported by the Israel Science Foundation F.I.R.S.T program under grant 1543/08 and grant 1421/08.
\end{acknowledgments}

\bibliographystyle{apsrev}
\bibliography{ucf2}

\begin{thebibliography}{29}
\bibitem[{rem({\natexlab{a}})}]{remark2}
\bibinfo{note}{It is reasonable to assume temperature independent spin-obit
  coupling if it is Rashba type. In addition, the dielectric constant does not
  change below 10K}.
\bibitem[{rem({\natexlab{b}})}]{remark}
\bibinfo{note}{The notable deviation of the dphasing rate at 0.4 K from the
  theoritacal prediction could be atributed to a supression of the
  superconducting fluctuations by the magnetic field at low temperatures}.
\end{thebibliography}
\end{document}